\documentclass[aps,prd,10pt,preprint,superscriptaddress,onecolumn,nofootinbib]{revtex4-1}
\usepackage{graphicx, epsfig} 
\usepackage{amsmath,amssymb,amsfonts,dsfont,mathrsfs,amsthm,mathtools}
\usepackage{bm} 
\usepackage{color}
\usepackage{float}
\usepackage[usenames]{xcolor}
\usepackage{hyperref}
\usepackage{siunitx}
\hypersetup{colorlinks=true,urlcolor=blue,linkcolor=magenta,citecolor=blue,filecolor=blue}
\usepackage[normalem]{ulem}
\usepackage{array}
\usepackage{hyperref}
\usepackage{booktabs}
\usepackage{blindtext}

\newcommand*{\Tr}{\operatorname{Tr}}

\newcommand{\diff}[1]{\text{d}#1}
\newcommand{\dd}{\text{d}}

\begin{document}

\title{Euclidean AdS wormholes and gravitational instantons in the Einstein-Skyrme theory}

\author{Fabrizio \surname{Canfora}}
\email{fabrizio.canfora@uss.cl}
\affiliation{Facultad de Ingenier\'ia, Arquitectura y Diseño, Universidad San Sebastian \\ General Lagos 1163, Valdivia 5110683, Chile}
\affiliation{Centro de Estudios Científicos (CECs), Casilla 1469, Valdivia, Chile}

\author{Crist\'obal \surname{Corral}}
\email{cristobal.corral@uai.cl}
\affiliation{Departamento de Ciencias, Facultad de Artes Liberales, Universidad Adolfo Ibáñez, Avenida Padre Hurtado 750, 2562340, Viña del Mar, Chile}

\author{Borja \surname{Diez}}
\email{bdiez@estudiantesunap.cl}
\affiliation{Instituto de Ciencias Exactas y Naturales, Universidad Arturo Prat, Avenida Playa Brava 3256, 1111346, Iquique, Chile}
\affiliation{Facultad de Ciencias, Universidad Arturo Prat, Avenida Arturo Prat Chac\'on 2120, 1110939, Iquique, Chile}

\begin{abstract}
Euclidean AdS wormholes provide a natural setup for studying the AdS/CFT correspondence with multiple boundaries. However, from a bottom-up perspective, they cannot be embedded in the four-dimensional Einstein-AdS-Maxwell theory if these boundaries have positive curvature. Nevertheless, Maldacena and Maoz showed that this obstruction could be circumvented by introducing merons in the four-dimensional Einstein-AdS-Yang-Mills theory. In this work, we show that Euclidean-AdS wormholes also exist in the four-dimensional Einstein-AdS-Skyrme theory, whose matter sector possesses a nontrivial baryonic charge. We compute its free energy and show that it does not depend on the integration constants whatsoever, resembling topological solitons. Additionally, we obtain its holographic stress tensor and show that it vanishes, allowing us to interpret this configuration as a holographic Bogomol'nyi–Prasad–Sommerfield (BPS) state. Other topologically nontrivial ground states in Einstein-Skyrme theory are found, such as gravitational instantons, representing the homotopically inequivalent vacua of the theory. We find that they develop Hawking-Page phase transitions above a critical temperature. Some of these solutions are periodic in Euclidean time, representing the gravitational analog of calorons in Yang-Mills theory. 
\end{abstract}

\maketitle

\section{Introduction\label{sec:intro}}

Euclidean AdS wormholes have attracted much interest since discovering the AdS/CFT correspondence~\cite{Maldacena:1997re,Witten:1998qj,Gubser:1998bc}. They connect two asymptotically AdS regions, providing a natural setting to study holography with more than one boundary~\cite{Witten:1999xp,Maldacena:2004rf,Bergshoeff:2005zf,Arkani-Hamed:2007cpn,Skenderis:2009ju}.
Their action can be lower than disconnected asymptotically AdS solutions, representing nontrivial saddle points in the gravitational partition function~\cite{VanRiet:2020pcn,Marolf:2021kjc,Chandra:2023rhx}. Indeed, their contribution to the latter can modify the factorization properties of the boundary correlators~\cite{Marolf:2021kjc,Betzios:2019rds,deBoer:2024mqg}, suggesting a possible modification of the standard picture of the AdS/CFT correspondence. These configurations are relevant when studying holographic entanglement entropy via the Ryu-Takayanagi proposal~\cite{Ryu:2006bv,Ryu:2006ef}, as shown in Refs.~\cite{Jensen:2014lua,Maxfield:2014kra,Balasubramanian:2014hda}. On the other hand, they can be used to describe black-hole microstates as they give a coarse-grained approximation to their energy levels via the gravitational constrained instanton method~\cite{Cotler:2021cqa,Cotler:2020lxj}. All these applications have motivated the quest for finding sensible theories where Euclidean AdS wormholes exist, such that novel aspects of the AdS/CFT correspondence with more than one conformal boundary can be explored. 

To construct these solutions, one must go beyond Einstein-AdS theory in vacuum. One possibility is to consider a top-bottom approach, such as supergravity theories~\cite{Maldacena:2004rf,Bergshoeff:2005zf,Arkani-Hamed:2007cpn,Hertog:2017owm,Anabalon:2020loe,Anabalon:2023kcp,Astesiano:2023iql} or in the presence of higher-curvature corrections coming from compactifications of string theory~\cite{Dotti:2006cp,Dotti:2007az,Dotti:2010bw}.\footnote{AdS wormholes were also found in Refs.~\cite{Blazquez-Salcedo:2020nsa,Chew:2021vxh} using phantom fields and in Ref.~\cite{Betzios:2024iay} with a Higgs field.} The stability of these configurations has been studied in Refs.~\cite{Correa:2008nq,Hertog:2018kbz,Anabalon:2019lzc,Hertog:2024nys}. Another possibility is considering a bottom-up approach. The simplest scenario, i.e. Einstein-Maxwell theory, does not support this class of solutions if the boundary has positive sectional curvature~\cite{Witten:1999xp,Cai:1999dqz}, producing instabilities on the dual conformal field theory. However, this can be circumvented by introducing nonlinear matter fields, such as merons~\cite{deAlfaro:1976qet} in Einstein-Yang-Mills theory, as Maldacena and Maoz showed in Ref.~\cite{Maldacena:2004rf}. Another possibility is to consider Einstein-AdS gravity coupled to the nonlinear sigma model, which can be obtained from the Einstein-Skyrme model by taking the limit of the Skyrme term going to zero~\cite{Ayon-Beato:2015eca,Ayon-Beato:2019tvu}. In those references, the authors obtained a Lorentzian AdS wormhole with two non-conformally flat boundaries due to the presence of a NUT parameter. In this work, we follow a different strategy: we focus on the Einstein-Skyrme model without neglecting the Skyrme term and show that an Euclidean AdS wormhole exists with two conformally flat boundaries of positive curvature.

The Skyrme model was first proposed in Ref.~\cite{SkyrmeOr} as a low-energy description for pions (see~\cite{SkyrmeRev} and references therein). In the $SU(2)$ case, the topologically trivial excitations of the theory represent pions while topological solitons correspond to nucleons; the latter are nowadays called \emph{Skyrmions} in honor of the revolutionary ideas of T.~Skyrme. In this context, the baryonic charge is associated with the third homotopy class of the $\mathfrak{su}(2)$-valued Skyrme field. Configurations with non-vanishing topological charge cannot decay into the trivial vacuum. Indeed, the original idea in Ref.~\cite{SkyrmeOr} was to supplement the nonlinear sigma model with an additional term to avoid the scaling argument of Derrick~\cite{Derrick}, which otherwise prevented the existence of topological solitons. An historical remark is in order: the Skyrme papers appeared before the Derrick paper so that the Derrick no-go theorem should be actually called Skyrme-Derrick no-go theorem. This proposal led to theoretical results in good agreement with the experiments~\cite{ANW}. The Skyrme term is the simplest possibility that circumvents the Derrick theorem and keeps, at the same time, the field equations covariant and of second order. This model has been widely studied in mathematical physics~\cite{SkyrmeRev}, and it plays a fundamental role both in nuclear and particle physics as it represents the low energy limit of QCD in the leading order of the 't Hooft large $N$ expansion~\cite{CaWi}.

Unlike what happens in Yang-Mills-Higgs theory, where the available BPS bounds and (anti-)self-duality conditions allow one to construct topological solitons analytically, the obvious BPS bound in the Skyrme theory cannot be saturated. This means that Skyrmions will produce a nontrivial backreaction to the Einstein-Skyrme Euclidean solutions, in contrast to what happens with (anti-)self-dual instantons since their stress-energy tensor vanishes. The Euclidean Einstein-Skyrme theory is particularly interesting from the viewpoint of gravitating topological solitons, providing access to an available nonperturbative window on interacting field theories~\cite{SkyrmeRev}. These smooth Euclidean solutions will be constructed generalizing the techniques in Refs.~\cite{CanforaP,Canfora:2013osa,Ayon-Beato:2015eca,Ayon-Beato:2019tvu} and we will see that they have a very rich phase diagram and very interesting geometric properties.

Another class of stable saddle points of the gravitational path integral are gravitational instantons. These configurations represent the ground state of homotopically inequivalent sectors of gravity theories, being stationary, regular, and Euclidean solutions characterized by topologically nontrivial indices of the Euler and Pontryagin classes. Some of them are (anti-)self-dual configurations in vacuum Einstein gravity, saturating a BPS bound. There are well-known examples of gravitational instantons in four dimensions~\cite{Hawking:1976jb,Eguchi:1978gw,Eguchi:1978xp,Gibbons:1978tef,Eguchi:1979yx,Lapedes:1980qw,Eguchi:1980jx,Chen:2011tc}, which have been extensively studied in gravity and in the low-energy limit of string theory~\cite{Khuri:1990mg,Johnson:1994ek,Burgess:1994kq,Kim:1996np,Beasley:2005iu,Dehghani:2005zm,Blumenhagen:2006xt,Dehghani:2006aa,Hendi:2008wq,Cagnazzo:2009zh,Palmer:2013haa,Carlevaro:2013vla,Flores-Alfonso:2018adn,Corral:2019leh,Corral:2022udb,Siahaan:2024ljt}. In the presence of self-gravitating Skyrmions, a generalization of the Eguchi-Hanson metric was found in Ref.~\cite{Canfora:2013osa} while Lorentzian solutions with nontrivial NUT charge were obtained in Ref.~\cite{Ayon-Beato:2015eca}. However, the backreaction of Skyrmions on the complex projective space $\mathbb{CP}^2$ and deformations thereof have not been studied. Here, we explore this possibility using a generalized hedgehog ansatz for a $SU(2)$-valued Skyrmion field and report the first solution of this class. 

The complex projective space $\mathbb{CP}^2$ is a compact, self-dual, and non-conformally flat Einstein-K\"ahler manifold that remains invariant under the action of the $SU(3)$ group. It has been considered a suitable internal manifold in Kaluza-Klein compactifications to generate non-Abelian interactions in supergravity~\cite{Duff:1986hr}. Their Euler characteristic and Hirzebruch signature are $\chi=3$ and $\tau=1$, respectively. Indeed, $\mathbb{CP}^2$ has been useful for studying the connection between super Yang-Mills theories and supersymmetric branes~\cite{Kim:2012tr} and for computing the partition function of the former in a dimensionally reduced theory over lens spaces~\cite{Lundin:2021zeb}. Moreover, the complex projective space provides a suitable background for computing the holomorphic anomaly of closed topological strings on Calabi-Yau manifolds by allowing one to perform its resurgent trans series nonperturbatively~\cite{Couso-Santamaria:2014iia}. 

In this work, we study topologically nontrivial ground states of the Einstein-Skyrme theory. To this end, we first focus on Euclidean-AdS wormholes and obtain a regular analytic solution supported by the self-gravitating Skyrmions. The partition function is obtained to first order in the saddle-point approximation. We show that the solution and the action thereof do not depend on any integration constant, resembling topological solitons. Moreover, we compute its holographic stress tensor and find that it vanishes in asymptotically AdS spacetimes. This allows us to interpret the solution as a BPS state from a conformal field theory viewpoint. Afterward, we focus on a different class of topologically nontrivial ground states by deforming $\mathbb{CP}^2$ due to the Skyrmions' backreaction. Locally, our solution remains invariant under the same isometry group. However, they are homotopically inequivalent, as indicated by their topological invariants. The solution represents two different thermodynamic states in the grand canonical ensemble described by two branches of the bolt radius as a function of the temperature. We find that the system develops Hawking-Page phase transitions between these thermal states, similar to what happens with AdS black holes~\cite{Hawking:1982dh}. Then, by considering a Skyrmionic field with nontrivial baryonic charge, we obtain a particular type of instanton that can be interpreted as the gravitational analog of calorons in Yang-Mills theory~\cite{Harrington:1978ve}; the latter are finite-temperature instantons periodic in Euclidean time. We show that the Skyrmions are crucial for rendering the solution regular, and we obtain specific constraints on the Skyrme term such that these gravitational calorons exist.

The manuscript is organized as follows: in Sec.~\ref{sec:EinsteinSkyrme}, we present the Einstein-Skyrme theory for the $SU(2)$ group and fix notation. Then, in Sec.~\ref{sec:EAdSWH}, we present the novel Euclidean AdS wormhole with self-gravitating Skyrmions and compute its partition function to first order in the saddle-point approximation. Section~\ref{sec:defCP2} is devoted to constructing new topologically nontrivial ground states of the theory by deforming the Fubini-Study metric of the complex projective space. We analyze the thermodynamics of this solution in the grand canonical ensemble where the temperature and topological charge are kept fixed and study the conditions under which Hawking-Page phase transitions occur. Then, in Sec.~\ref{sec:calorons}, we construct the gravitational caloron with self-gravitating Skyrmions with a nonvanishing baryonic charge. Finally, we discuss and summarize our main results in Sec.~\ref{sec:discussion}.

\section{The Einstein-Skyrme theory\label{sec:EinsteinSkyrme}}

In this section, we review the Einstein-Skyrme theory and fix notation. We will work mainly in Euclidean signature unless stated otherwise. The action principle describing the dynamics of the Einstein-Skyrme theory is given by 
\begin{align}\notag
    I_E[g_{\mu\nu},B_\mu] &= \kappa\int_{\mathcal{M}}\diff{^4x}\sqrt{|g|}(R-2\Lambda) + \frac{K}{4}\int_{\mathcal{M}}\diff{^4x}\sqrt{|g|}\Tr\left(B^\mu B_\mu +\frac{\lambda}{8}F_{\mu\nu }F^{\mu\nu}\right)\\
    &+ 2\kappa\int_{\partial \mathcal{M}}\diff{^3x}\sqrt{|h|}\,\mathcal{K} + I_{\rm CT} := I_{\rm bulk} + I_{\rm GHY} + I_{\rm CT}\,, \label{action}
\end{align}
where $\kappa=(16\pi G)^{-1}$ denotes the gravitational constant, $B_\mu = U^{-1}\nabla_\mu U$ is a $\mathfrak{su}(2)$-valued vector field where $U\in SU(2)$, and $F_{\mu\nu}=\left[B_\mu,B_\nu \right]$. The matter sector is parametrized by two different coupling constants, i.e., $K$ and $\lambda$. As already emphasized, Skyrme included the term proportional to $\lambda$ to allow the existence of static topological solitons describing baryons. The first boundary term, i.e. $I_{\rm GHY}$, is the Gibbons-Hawking-York term, which is required for defining a well-posed variational principle with Dirichlet boundary conditions for the induced metric $h_{ab}$. The latter can be read off from radial decomposition in Gauss-normal coordinates, i.e. 
\begin{align}\label{gaussnormal}
    \diff{s^2} = N^2(r)\diff{r^2} + h_{ab}(r,x)\diff{x^a}\diff{x^b}\,,
\end{align}
where $N^2(r)$ is the lapse function, $\{x^a\}$ denote boundary coordinates, and $h_{ab}(r,x)$ is the codimension-1 induced boundary metric. Hereon, Latin characters from the beginning of the alphabet represent boundary indices. Using this foliation, the extrinsic curvature can be expressed as $\mathcal{K}_{ab}=-\tfrac{1}{2N}\partial_r h_{ab}$. Additionally, $I_{\rm CT}$ are intrinsic boundary counterterms needed to render the action and variations thereof finite on asymptotically locally AdS spacetimes.

The field equations are obtained by performing arbitrary variations with respect to the metric and $\mathfrak{su}(2)$-valued Skyrme field, giving
\begin{subequations}\label{eom}
\begin{align}\label{eomg}
    R_{\mu\nu} - \frac{1}{2}g_{\mu\nu}R + \Lambda g_{\mu\nu} &= \frac{1}{2\kappa}T_{\mu\nu}\,,\\
    \label{eomB}
    \nabla_\mu\left(B^\mu + \frac{\lambda}{4}\left[B_\nu,F^{\mu\nu} \right] \right) &= 0\,,
\end{align}
\end{subequations}
respectively, with the stress-energy tensor being defined as
\begin{align}\label{Tmunu}
    T_{\mu\nu }=-\frac{K}{2}\Tr\left[\left(B_\mu B_\nu -\frac{1}{2}g_{\mu\nu }B^\alpha B_\alpha \right)+\frac{\lambda}{4}\left(F_{\mu\alpha }F_\nu^{~\alpha }-\frac{1}{4}g_{\mu\nu }F_{\alpha\beta }F^{\alpha\beta}\right)\right].
\end{align}
The trace of the latter is quadratic in $B_\mu$, as the contribution coming from the Skyrme term vanishes, that is, $g^{\mu\nu}T_{\mu\nu}= \tfrac{K}{2}\Tr[B_\mu B^\mu]$.

In this work, we use $t_i=-\tfrac{i}{2}\tau_i$ as the anti-Hermitian generators of $SU(2)$, where $\tau_i$ are the Pauli matrices. Hereon, we use Latin characters of the middle of the alphabet to denote $SU(2)$ indices. Using these definitions, we have 
\begin{align}
    \left[t_i,t_j\right] = \epsilon_{ijk}t^k \;\;\;\;\; \mbox{and} \;\;\;\;\; t_it_j = -\frac{1}{4}\delta_{ij}\mathbb{I} + \frac{1}{2}\epsilon_{ijk}t^k\,.
\end{align}
The last relation implies that $\Tr(t_it_j)=-\tfrac{1}{2}\delta_{ij}$ and $\Tr(t_it_jt_k)=-\tfrac{1}{4}\epsilon_{ijk}$. Additionally, it is convenient to introduce the left-invariant forms of $SU(2)$, which can be parametrized in terms of the Euler angles as
\begin{subequations}\label{LIforms}
\begin{align}
\sigma_1&=\cos\psi\diff{\vartheta}+\sin\vartheta\sin\psi\diff{\varphi}\,,\\
  \sigma_2&=-\sin\psi\diff{\vartheta}+\sin\vartheta\cos\psi\diff{\varphi},\\
  \sigma_3&=\diff{\psi}+\cos\vartheta\diff{\varphi}\,,
\end{align}    
\end{subequations}
and they satisfy the Maurer-Cartan equation $\diff{\sigma_i}+\frac{1}{2}\epsilon_{ijk}\sigma^j\wedge \sigma^k=0$, where $SU(2)$ indices are raised and lowered by the Cartan-Killing metric of such a group, while $\wedge$ denotes the wedge product of differential forms. 

Different topological invariants characterize the nontrivial topology of the solutions presented in this work. For instance, in our conventions, the baryonic charge is defined in terms of the boundary integral 
\begin{align}\label{topologicalcharge}
    Q_B = -\frac{1}{24\pi^2}\int_{\partial\mathcal{M}}\Tr\left(\,B\wedge B\wedge B \right)\,,
\end{align}    
where $B=B_\mu\diff{x^\mu}$ and $\partial{\mathcal{M}}$ is a codimension-1 hypersurface. Additionally, the Euler characteristic $\chi(\mathcal{M})$ can be obtained directly from the Gauss-Bonnet theorem, that is,
\begin{align}\label{Euler}
\int_{\mathcal{M}}\diff{^4x}\sqrt{|g|}\,\mathcal{G} = 32\pi^2\chi(\mathcal{M}) + \int_{\partial\mathcal{M}}\diff{^3x}\sqrt{|h|}\,\mathcal{B}\,,
\end{align}
where the Gauss-Bonnet term and the Chern form are respectively given by
\begin{subequations}\label{GB}
    \begin{align}
   \mathcal{G} & = R^2 - 4R^\mu_\nu R^\nu_\mu + R^{\mu\nu}_{\lambda\rho}R^{\lambda\rho}_{\mu\nu} \,,\\
    \mathcal{B} &= -4\delta^{abc}_{def}\mathcal{K}^d_a\left(\frac{1}{2}\mathcal{R}^{ef}_{bc}(h) - \frac{1}{3}\mathcal{K}^e_b \mathcal{K}^f_c \right)\,,
\end{align}
\end{subequations}
with $\mathcal{R}^{ab}_{cd}(h)$ being intrinsic curvature. In Einstein-AdS gravity, the two terms in Eq.~\eqref{GB} can be equivalently used as counterterms in spacetimes with AdS asymptotic, and they coincide with standard holographic renormalization to relevant order in the Fefferman-Graham expansion~\cite{Aros:1999id,Aros:1999kt,Olea:2005gb,Miskovic:2009bm}. Here, however, we will consider intrinsic boundary counterterms as we are not dealing with Einstein spaces. Indeed, the contribution of Skyrmions becomes nontrivial at the boundary, and they can be used to describe confinement or deconfinement phases of (quasi-)particles at strong coupling in the dual super Yang-Mills theories~\cite{Cartwright:2020yoc,Cartwright:2021tcy}.

\section{Euclidean AdS wormhole with nontrivial Baryonic charge\label{sec:EAdSWH}}

To find a Euclidean wormhole with a negative cosmological constant, $\Lambda=-3/\ell^2$, where $\ell$ denotes the AdS radius, we assume the following ans\"atze for the metric and Skyrme field
\begin{align}\label{ansatzWH}
\diff{s^2} = \frac{\diff{r^2}}{f(r)} + (r^2+r_0^2)\,\diff{\Omega_{(3)}^2} \;\;\;\;\; \mbox{and} \;\;\;\;\;  B_\mu = \sigma_\mu^i t_i \,.
\end{align}
Here, $r_0\in\mathbb{R}/\{0\}$ denotes the wormhole throat, $\diff{\Omega_{(3)}^2}=\tfrac{1}{4}(\sigma_1^2+\sigma_2^2+\sigma_3^2)$ is the metric of the round three-sphere, and $\sigma^i=\sigma^i_\mu\diff{x^\mu}$ defines the components of the left-invariant forms of $SU(2)$. Using Eq.~\eqref{ansatzWH} for the metric and the matter content, one can check that the Skyrme equations~\eqref{eomB} are automatically satisfied. The Skyrmion's field strength can be read directly from its definition, giving
\begin{equation}
    F_{\mu\nu }=\epsilon_{ijk}\sigma_\mu^{i}\sigma_\nu^j t^k\,.
\end{equation}
This configuration has a nontrivial baryonic charge, whose value can be obtained by evaluating Eq.~\eqref{topologicalcharge}, giving~\cite{Ayon-Beato:2015eca}
\begin{align}\label{topologicalcharge1}
    Q_B = 1\,.
\end{align}

The Euclidean Einstein-Skyrme system admits solutions for arbitrary values of the coupling constants. Although one can show that it possesses a closed-form solution if $K\neq4\kappa$, here we focus on the case $K=4\kappa$ as the analysis of the regularity of the partition function is especially neat.\footnote{This analysis can be generalized to arbitrary values of the matter coupling using the holographic renormalization prescription~\cite{Henningson:1998gx,deHaro:2000vlm,Bianchi:2001kw,Skenderis:2002wp}.} It is worth emphasizing that this limit cannot be taken smoothly at the level of the solutions of Sec.~\ref{sec:calorons}, where a positive or zero cosmological constant is considered. Then, inserting the Skyrme solution in Eq.~\eqref{ansatzWH} into the field equations~\eqref{eomg}, the systems reduces to
\begin{subequations}
    \begin{align}\label{WHeq1}
        \frac{3r^2f}{(r+r_0^2)^2}+\frac{3\lambda}{(r^2+r_0^2)^2}-\frac{3}{\ell^2}&=0\,,\\
        \frac{rf'}{r^2+r_0^2}+\frac{f(r^2+2r_0^2)}{(r^2+r_0^2)^2}-\frac{\lambda}{(r^2+r_0^2)^2}-\frac{3}{\ell^2}&=0\,, \label{WHeq2}
    \end{align}
\end{subequations}
where $f=f(r)$ and prime denotes differentiation with respect to the radial coordinate $r$. The above system can be solved analytically by first integrating Eq.~\eqref{WHeq2}. This leads to a one-parameter family of solutions characterized by a single integration constant. Then, replacing that result into Eq.~\eqref{WHeq1}, one concludes that the integration constant has to vanish. Then, the system is solved by the metric function
\begin{align}\label{fwh0}
f(r)=\frac{r^2}{\ell ^2} + \frac{2r_0^2}{\ell^2} + \frac{r_0^4 - \ell^2\lambda}{\ell^2\,r^2}   \,.
\end{align}
The singularity at $r=0$ can be removed by demanding the last term to vanish. This fixes the wormhole throat in terms of the Skyrme coupling as $r_0=(\ell^2\lambda)^{1/4}$. Indeed, reality conditions on $r_0$ impose no further constraints as $\lambda>0$; this is required by consistency with experimental data (see Ref.~\cite{ANW}). Inserting this value into Eq.~\eqref{fwh0}, we find that the metric function that solves the field equations becomes
\begin{align}\label{fsolWH}
    f(r) = \frac{r^2}{\ell^2} + \frac{2\sqrt{\lambda}}{\ell}\,.
\end{align}
This is a regular solution supported by nontrivial Skyrmions that represents an Euclidean AdS wormhole connecting two conformal boundaries with the same AdS radius. It has a $\mathbb{Z}_2$ symmetry relating two three-spheres at a given radial coordinate. Additionally, it lacks any integration constant while having a nontrivial baryonic charge [cf. Eq.~\eqref{topologicalcharge1}]. This allows us to interpret this horizonless regular solution as a gravitational topological soliton. As we will see next, it has a finite action that depends only on the coupling constants, suggesting that it represents a topologically nontrivial ground state of the theory.  



\subsection{Euclidean action and topological invariants}

To compute the global properties of the solution, we need to separate the boundary into two pieces, i.e. $\partial\mathcal{M}=\partial\mathcal{M}_+\cup \partial \mathcal{M}_-$, located at $r\to\pm\infty$, respectively. In the case of the Euler characteristic, this implies
\begin{align}\label{Euler-char}
\int_{\mathcal{M}}\diff{^4x}\sqrt{|g|}\,\mathcal{G} = 32\pi^2\chi(\mathcal{M}) + \int_{\partial\mathcal{M}_+}\diff{^3x}\sqrt{|h_+|}\,\mathcal{B}_+ - \int_{\partial\mathcal{M}_-}\diff{^3x}\sqrt{|h_-|}\,\mathcal{B}_-\,,
\end{align}
where $h_\pm$ are the determinant of the induced metric at the boundaries $\partial\mathcal{M}_\pm$, respectively, and $\mathcal{B}_\pm$ is the Chern form evaluated at those codimension-1 hypersurfaces. In the case of the Euclidean AdS wormhole, explicit computation of these integrals yield
\begin{align}
    \int_\mathcal{M} \dd^4x\sqrt{|g|}\,\mathcal{G}&=-\frac{16\pi^2r\sqrt{r^2+2r_0^2}}{\ell^3(r^2+r_0^2)^\frac{3}{2}}\bigg[r_0^2(3\ell^2-2r^2)+r^2(3\ell^2-r^2)\bigg]_{r=r_-}^{r=r_+}\,, \\
\int_{\partial\mathcal{M}_\pm}\dd^3x\sqrt{|h_\pm|}\,\mathcal{B}_\pm&=\frac{16\pi^2r_\pm\sqrt{r_\pm^2+2r_0^2}}{\ell^3(r_\pm^2+r_0^2)^\frac{3}{2}}\left[r_0^2(3\ell^2-2r_\pm^2)+r_\pm^2(3\ell^2-r_\pm^2)\right]\,.
\end{align}
Notice that the bulk integral cancels the boundary terms before removing the regulators $r_\pm$. Therefore, the Euler characteristic of the Euclidean AdS wormhole gives $\chi(\mathcal{M})=0$.

The partition function, on the other hand, can be computed to first order in the saddle-point approximation through $\ln\mathcal{Z}\approx-I_E$, where $I_E$ is the Euclidean on-shell action. Its value, however, is generically divergent. To renormalize it, we introduce intrinsic boundary counterterms of the form
\begin{align}
    I_{\rm CT} = 2\kappa\int_{\partial\mathcal{M}}\diff{^3x}\sqrt{|h|}\left(z_0 + z_1 \, \mathcal{R} + z_2 \Tr\left[B^a B_a\right]+z_3 \Tr\left[F_{ab}F^{ab}\right] + \ldots \right)\,, \label{ICT}
\end{align}
where $\mathcal{R}=\mathcal{R}(h)$ is the intrinsic curvature scalar of the induced metric $h_{ab}$; the latter is used to raise and lower the boundary indices. Additionally, the couplings $z_i$ are fixed by demanding regularity. As the solution has two boundaries and the on-shell action is odd in the radial coordinate, it suffices to expand the action as $r\to\infty$, fix the couplings to cancel divergencies, and then sum up the contribution coming $r\to-\infty$. Following this prescription, we find that the behavior of the Euclidean on-shell action is
\begin{align}\notag 
    I_E &=  -8\kappa\pi^2\Bigg[\left(\frac{2+z_0\ell}{2\ell}\right)r^3 + \frac{3}{4}\left[4\ell(z_1 - z_2) + (2 + \ell z_0)\sqrt{\lambda}\right]\, r  \\ 
    &+ \lambda^{3/4}\sqrt{2\ell}\;i\,\text{F}\left(\frac{i\,r}{(\ell^2\lambda)^{1/4}},\frac{\sqrt{2}}{2}\right)\Bigg] + \mathcal{O}(r^{-1})\,,
\end{align}
as $r\to\infty$. Here, $\text{F}(z,k)$ is the incomplete elliptic function of the first kind, which is odd in the first argument, i.e. $\text{F}(-z,k)=-\text{F}(z,k)$. Additionally, its asymptotic limit is
\begin{align}
    \lim_{x\to\pm\infty}\,i\,\text{F}(i\,x,k) = \mp\,\text{K}(1-k)\,,
\end{align}
where $\text{K}(x)$ is the complete elliptic function of the first kind. Under these considerations, renormalization demands
\begin{align}\label{zsol}
    z_0 = -\frac{2}{\ell} \;\;\;\;\; \mbox{and} \;\;\;\;\; z_1 = z_2\,,
\end{align}
without imposing any constraint on the value of $z_3$. This is related to the fact that the falloff of the fourth term in Eq.~\eqref{ICT} contributes neither to the divergent nor the finite part of the action. Then, the renormalized on-shell action for the Euclidean AdS wormhole is 
\begin{align}
    I_{\rm WM}^{\rm (ren)} = 16\pi^2\kappa\lambda^{3/4}\sqrt{2\ell}\;\text{K}_*\,,
\end{align}
where the numerical value of the complete elliptic function of the first kind appearing in the last equation is $\text{K}_*:= \text{K}\left(1-\frac{\sqrt{2}}{2}\right) \approx 1.60621$. As anticipated, neither the Euclidean AdS wormhole nor its partition function depends on any integration constant whatsoever. This suggests that it can be interpreted as a topological soliton, representing a topologically nontrivial ground state of the theory.

The partition function of the Euclidean AdS wormhole can be compared with that of the global Euclidean AdS space. The latter represents a disconnected solution possessing a single conformal boundary. In order to compare the Euclidean AdS ground state with the two-boundary AdS wormhole, we consider two copies of the former, such that each of their boundaries is identified with those of the latter (see Ref.~\cite{Marolf:2021kjc} for details). In the hyperspherical coordinates used in Eq.~\eqref{ansatzWH}, this ground state is obtained by setting $r_0=0$ and $f(r)=1+r^2/\ell^2$. Neglecting the Skyrmions, the renormalized action is obtained by setting $z_0 = -2/\ell$ and $z_1 = -\ell/2$ in Eq.~\eqref{ICT}, giving
\begin{align}
    I_{\rm AdS}^{\rm (ren)} = 8\pi^2\kappa\ell^2 \,.
\end{align}
The difference between the actions of these ground states, i.e. $\Delta I_E = I_{\rm WM}^{\rm (ren)} - 2I_{\rm AdS}^{\rm (ren)}$, allow us to determine which of them dominates the path integral. Thus, if $\Delta I_{\rm E}<0$, the Euclidean AdS wormhole will dominate over the disconnected solution. This occurs whenever the Skyrme parameter lies in the range $0<\lambda<\lambda_*$, where 
\begin{align}
    \lambda_* = \frac{2^{2/3}\ell^2}{(2\text{K}_*)^{4/3}}\,.
\end{align}
Thus, we conclude that there is a range in parameter space where the Euclidean AdS soliton dominates over the global Euclidean AdS solution, representing the topologically nontrivial ground state of the theory. 

\subsection{Renormalized holographic stress-energy tensor and non-Abelian current}

The renormalized quasilocal stress-energy tensor for Einstein-Skyrme theory on asymptotically AdS spacetimes can be obtained by varying the action principle~\eqref{action} with respect to the induced metric $h_{ab}$, where the counterterm action is given in Eq.~\eqref{ICT}. In the case of Einstein-AdS gravity, it was obtained by Balasubramanian-Kraus in Ref.~\cite{Balasubramanian:1999re}. In the presence of the Skyrme terms, the renormalized stress-energy tensor is
\begin{align}\notag
    \tau_{ab} &= -2\kappa\bigg(\mathcal{K}_{ab} - h_{ab}\mathcal{K} - z_0\, h_{ab} + z_1\left(\mathcal{R}_{ab} - \frac{1}{2}h_{ab}\mathcal{R} \right) + z_2\Tr\left[B_aB_b - \frac{1}{2}h_{ab}B_c B^c \right] \\
    &\qquad\qquad + z_3\Tr\left[F_{ac}F_{b}{}^{c} - \frac{1}{4}h_{ab}F_{cd}F^{cd} \right] + \ldots \bigg)\,,
\end{align}
where $z_0$ and $z_1$ are given in Eq.~\eqref{zsol}, while $z_3$ remains free as the last terms are subleading towards the asymptotic boundary. Asymptotic charges associated with a Killing vector field $\xi=\xi^\mu\partial_\mu$ are defined by the codimension-2 boundary integral
\begin{align}\label{BKcharges}
     Q[\xi] = \int_{\Sigma_\infty}\diff{^2x}\sqrt{|\sigma|}\, \tau^a_b\, \xi^b\,u_a\,,
\end{align}
where $\Sigma_\infty$ is a codimension-2 asymptotic boundary with $u=u^\mu\partial_\mu$ being its unit-normal and $\sigma$ is the determinant of its induced metric.  

According to the AdS/CFT correspondence~\cite{Maldacena:1997re,Witten:1998qj,Gubser:1998bc}, the holographic stress tensor $\mathcal{T}_{ab}$ can be obtained from the renormalized quasilocal stress-energy tensor by taking the limit
\begin{align}\label{holographicTmunu}
    \langle\, \mathcal{T}_{ab}\, \rangle = \lim_{r\to\infty}\,r\,\tau_{ab} \,.
\end{align}
For the Euclidean AdS wormhole solution sourced by Skyrmions, we find that the components of the renormalized quasilocal stress-energy tensor are
\begin{subequations}
\begin{align}
    \tau_{\psi\psi} &=  \frac{\kappa(4z_3-\ell\lambda )}{2r^2} - \frac{\kappa \ell\sqrt{\lambda}(4z_3-\ell\lambda )}{2r^4} + \mathcal{O}(r^{-6})\,, \\
    \tau_{\psi\varphi} &=  \frac{\kappa(4z_3-\ell\lambda )\cos\vartheta}{2r^2} - \frac{\kappa \ell\sqrt{\lambda}(4z_3-\ell\lambda )\cos\vartheta}{2r^4} + \mathcal{O}(r^{-6})\,, 
\end{align}    
\end{subequations}
and $\tau_{\vartheta\vartheta} = \tau_{\varphi\varphi}=\tau_{\psi\psi}$. Thus, from the conformal field theory viewpoint, the AdS wormhole solution with Skyrme source has a vanishing holographic stress tensor, as it can be checked from Eq.~\eqref{holographicTmunu}. Then, all the conserved charges~\eqref{BKcharges} will be zero. However, it has a nontrivial topological charge as shown in Eqs.~\eqref{topologicalcharge} and~\eqref{topologicalcharge1}. This exactly happens with configurations that saturate the BPS bound. Therefore, we conclude that this solution represents a holographic BPS state of the dual CFT. 

The 1-point function of the holographic $SU(2)$ current can be obtained from the boundary value of the Skyrmion field. The Fefferman-Graham expansion of the latter is given by
\begin{align}
    B_a = B^{(0)}_a + \frac{1}{r}\,B^{(1)}_a + \frac{1}{r^2}\,B^{(2)}_a + ... \,,
\end{align}
where $B_a^{(I)} = B_a^{(I)\,i}\,t_i$, with $I\in\mathbb{Z}_{\geq0}$. From the Euclidean AdS wormhole solution with Skyrme source, we notice that $B_a^{(0)} = \sigma_a^i\,t_i$ while $B_a^{(m)}=0$ if $m\geq1$. Then, the AdS/CFT dictionary allows one to obtain the vacuum expectation value of the dual $SU(2)$ current and its corresponding source; they are
\begin{align}
    \langle\, \mathcal{J}_a \,\rangle = \lim_{r\to\infty} r B_a = B_a^{(1)} = 0 \;\;\;\;\; \mbox{and} \;\;\;\;\; \mathcal{A}_a^{\rm ext} = \lim_{r\to\infty} B_a  = B_a^{(0)} = \sigma_a^i\, t_i\,,
\end{align}
respectively. The vanishing of the former can be interpreted as a confining phase in the dual field theory produced by a topologically nontrivial source with a winding number equal to 1. This is similar to what was obtained in Refs.~\cite{Cartwright:2020yoc,Cartwright:2021tcy} for the Einstein-Skyrme theory in five dimensions.

\section{Gravitational instantons as deformations of $\mathbb{CP}^2$\label{sec:defCP2}}

As mentioned in the introduction, a different topologically nontrivial ground state of the Einstein-Skyrme theory are gravitational instantons. In this section, we consider $\mathbb{CP}^2$ and deformations thereof due to the backreaction of self-gravitating Skyrmions and study their Hawking-Page phase transitions. 

In Ref.~\cite{Canfora:2013osa}, the hedgehog ansatz was generalized to construct nonlinearly charged black holes, Taub-NUT, and Eguchi-Hanson spacetimes in the presence of self-gravitating Skyrmions. To do so, the authors considered an element of $SU(2)$ parametrized by
\begin{align}\label{hedgehog}
    U = \cos\alpha\,\mathbb{I} + 2\,n^i\, t_i\,\sin\alpha := U_0\,\mathbb{I} + U^i\, t_i\,,
\end{align}
where $\alpha=\alpha(x)$ is a scalar function and $n^i = (\sin\vartheta\cos\varphi,\sin\vartheta\sin\varphi,\cos\vartheta)$ is a unit-normal vector of $SU(2)$ satisfying $n_in^i=1$ and $n_i\nabla_\mu n^i=0$. Then, it is direct to see that $\mathfrak{su}(2)$-valued vector field can be written as
\begin{align}\label{Bmuhedgehog}
    B_\mu = U^{-1}\nabla_\mu U = \left(U_0\nabla_\mu U_i - U_i\nabla_\mu U_0 - \frac{1}{2}\epsilon_{ijk}U^j\nabla_\mu U^k \right)t^i\,.
\end{align}
Henceforth, we will use this ansatz and focus on the case when $\alpha=\pi/2+n\pi$ with $n\in\mathbb{Z}$. 

To solve the Einstein-Skyrme field equations, in addition to the ansatz in Eq.~\eqref{Bmuhedgehog}, we assume a Riemannian line element given by
\begin{equation}\label{deformedCP2}
  \dd s^2=\frac{\dd r^2}{f(r)(1+ar^2)^2}+f(r)h(r)\frac{r^2}{4}\frac{\sigma_3^2}{(1+ar^2)^2}+\frac{r^2}{4}\frac{(\sigma_1^2+\sigma_2^2)}{(1+ar^2)}\,,
\end{equation}
where $a$ is a constant with units of $\mbox{length}^{-2}$. Notice that, if $f(r)=h(r)=1$, it reduces to the Fubini-Study metric of $\mathbb{CP}^2$. The latter is a regular compact space that remains invariant under the action of the $SU(3)$ isometry group and it belongs to the family of gravitational instantons of general relativity in vacuum~\cite{Eguchi:1980jx}. Moreover, it is an Einstein space with $R_{\mu\nu}=\Lambda g_{\mu\nu}$, if $a=\Lambda/6$, possessing Euler characteristic and Hirzebruch signature $\chi=3$ and $\tau=1$, respectively. Indeed, one can check that the Skyrme equation is satisfied automatically when the metric and Skyrmion ans\"atze in Eqs.~\eqref{deformedCP2} and~\eqref{Bmuhedgehog} are used. This, of course, includes the particular case of Skyrmions on the $\mathbb{CP}^2$ background, but it allows us to study deformations of it by introducing the Skyrmions' backreaction. Then, taking into account Eq.~\eqref{Bmuhedgehog}, we find that the Einstein-Skyrme equations are solved by $h(r)=1$ and
\begin{align}\notag
    f(r) &= \frac{p(ar^2+1)^3}{r^4} + \frac{\Lambda[3ar^2(ar^2+1)+1]}{6a^3 r^4} + \frac{2K\lambda(ar^2+1)^2}{a\kappa r^4} \\ &\quad + \frac{(K-2\kappa)(ar^2+1)(2ar^2+1)}{2a^2\kappa r^4}\,, \label{soldeformedCP2-f-1/f}
\end{align}
where $p$ is an integration constant and $a\neq0$. This is a one-parameter extension of the Eguchi-Hanson solution with self-gravitating Skyrmions found in Ref.~\cite{Canfora:2013osa}. Indeed, Eq.~\eqref{soldeformedCP2-f-1/f} is not continuously connected to the latter in the $a\to0$ limit, as the Kretschmann invariant behaves as $\mathcal{O}(a^{-6})$ in the presence of $\Lambda$ and as $\mathcal{O}(a^{-4})$ in the absence thereof. Additionally, if $K=0$, that is, in the vanishing Skyrmions' limit, one can check that this solution becomes the Fubini-Study metric of $\mathbb{CP}^2$ if the simultaneous limit $p\to0$ and $a\to\Lambda/6$ is taken.

There is always a region in parameter space where the metric has a single two-dimensional set of fixed points ---a bolt--- at the locus $r=r_b$ defined by the largest positive root of the polynomial $f(r_b)=0$. The absence of conical singularities, in turn, implies that $\psi\sim\psi+\beta$, where 
\begin{align}\label{betaf-1/f}
    \beta = \frac{8\pi\kappa r_b^2}{\left|4K\lambda(ar_b^2+1)+2r_b^2(K-2\kappa) + \frac{\Lambda\kappa r_b^4}{ar_b^2+1}\right|}\,,
\end{align}
is the period of Euclidean time, which can be interpreted as the inverse of the Hawking temperature. By solving $r_b$ in terms of the temperature, we obtain
\begin{equation}\label{rhT}
    r_{\pm}^2(T)=\frac{-K(4a\lambda+1)+2\kappa(1-2\pi T)\pm \sqrt{(4\pi T\kappa+K-2\kappa)^2-4K\kappa\Lambda\lambda}}{K(4a^2\lambda+2a)+\kappa(8\pi Ta+\Lambda-4a)}\,.
\end{equation}
These represent two different thermodynamic states, whose free energies can be compared to see whether Hawking-Page phase transitions exist. 

To analyze the boundary metric of this solution, we consider radial foliation defined in Eq.~\eqref{gaussnormal}. In these coordinates, the boundary is located at $r\to\infty$ and it can be written in terms of the left-invariant forms of $SU(2)$ as
\begin{align}\label{dsbdy}
    \diff{s^2_{\rm bdy}} 
    = \frac{1}{4a}\left(\sigma_1^2 + \sigma_2^2 + \mu^2\, \sigma_3^2 \right)\,,
\end{align}
where $\mu^2 = a^2p$. Then, it becomes clear that the boundary metric is a squashed three-sphere, whose squashing parameter is controlled by the integration constant $p$, while its radius is $a^{-1/2}$. This codimension-1 hypersurface is nonconformally flat, as their boundary Cotton tensor, i.e. $\mathcal{C}^{ab}=\varepsilon^{acd}\mathcal{D}_d\left(\mathcal{R}^b_c - \frac{1}{4}\mathcal{R}\delta^b_c \right)$, is nonvanishing as long as $a^2p\neq1$; here, calligraphic objects are intrinsic quantities constructed out of the induced metric $h_{ab}$. Asymptotically locally AdS spacetimes with a conformal boundary of the form~\eqref{dsbdy} have been relevant for describing holographic fluids with nontrivial vorticity~\cite{Leigh:2011au,Leigh:2012jv,Caldarelli:2012cm,Mukhopadhyay:2013gja}, for computing holographic correlators in $\mathcal{N}=4$ supersymmetric field theories on squashed spheres~\cite{Bobev:2016sap,Bobev:2017asb}, and for unveiling universal aspects of their partition function~\cite{Bueno:2018yzo}.

\begin{figure}[h!]
    \centering
    \includegraphics[width=0.7\linewidth]{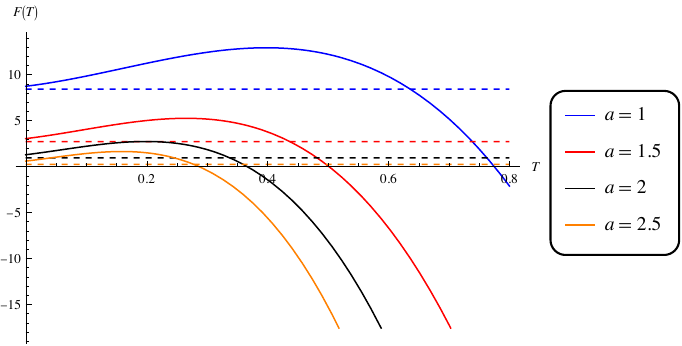}
    \caption{The free energy as a function of the temperature for the two branches in Eq.~\eqref{rhT} for different values of $a$. The solid line represents the free energy of the branch $r^2_{+}(T)$, while the dashed line stands for $r^2_{-}(T)$. Then, it is clear that, above the critical temperature, the branch $r^2_+(T)$ dominates over $r^2_-(T)$. This represents a Hawking-Page phase transition between these gravitational instantons.}
    \label{fig:plot-Free-Energy}
\end{figure}

The Euclidean on-shell action can be obtained by computing each term of Eq.~\eqref{action} separately. In the case $h(r)=1$, direct computation of the bulk action yields
\begin{align}
   I_{\rm bulk} &= -2\pi\beta K\lambda\ln\left(\frac{ar_b^2}{ar_b^2+1}\right)-\frac{\pi\beta\Lambda\kappa(2ar_b^2+1)}{4a^2(ar_b^2+1)^2} \,,
\end{align}    
which is finite towards the asymptotic boundary. Notice that the limit $\Lambda\to0$ is smooth in this case. Therefore, there is no need to add intrinsic boundary counterterms as in asymptotically locally AdS spacetimes, since the variational principle is well-posed by virtue of the Gibbons-Hawking-York term only, whose on-shell value is
\begin{align}
    I_{\rm GHY}&=\frac{\pi\beta K(2a\lambda+1)}{a}+\frac{\pi\kappa\beta(\Lambda-4a-2a^3p)}{2a^2}\,.
\end{align}
Then, summing up bulk and boundary contributions, we arrive at the Euclidean on-shell action for this configuration, that is,
\begin{align}\notag
    I_E &= 2\pi\beta K\lambda\left[\frac{ar_b^2+2}{ar_b^2+1}-\ln\left(\frac{ar_b^2}{ar_b^2+1} \right)\right] + \frac{\pi\beta(K-2\kappa)(2a^2r_b^4+6ar_b^2+3)}{2a(ar_b^2+1)^2}\\
    &+\frac{\pi\beta\kappa\Lambda(6a^3r_b^6+18a^2r_b^4+15ar_b^2+5)}{12a^2(ar_b^2+1)^3}\,.   
\end{align}
In this case, the logarithmic term appears exclusively due to the presence of the Skyrme term, and it vanishes in the $r_b\to\infty$ limit. The free energy can be obtained from $\mathcal{F}=\beta^{-1}I_E$. Additionally, for different values $a$, the free energy as a function of the temperature is given in Figure~\ref{fig:plot-Free-Energy}. One can see that a Hawking-Page phase transition occurs at a critical temperature $T=T_c$, which can be obtained from $\Delta I^\pm_E(T_c) = 0$, where $\Delta I^\pm_E = I_E^+(T) - I_E^-(T)$ is the difference between the Euclidean actions of the branches $r_\pm^2(T)$, respectively. Indeed, the critical temperature decreases as $a$ increases; this can be seen from the behavior of the free energies in Figure~\ref{fig:plot-Free-Energy}.

Let us focus on the case $h(r)\neq1$. Inserting the metric~\eqref{deformedCP2} and the hedgehog ansatz~\eqref{Bmuhedgehog} into the Einstein-Skyrme field equation, we find that they can be written as a first-order system, that is,
\begin{subequations}
    \begin{align}
    \frac{\diff{h}}{\diff{r}}&=\frac{2h(h-1)}{r(ar^2+1)}\,,\\
    \label{dfdr}
    \frac{\diff{f}}{\diff{r}}&=-\frac{4(ar^2+1)K\lambda}{\kappa r^3}-\frac{f(3h+1-2ar^2)}{r(ar^2+1)}-\frac{2r^2Ka+\Lambda r^2\kappa-4a\kappa r^2+2K-4\kappa}{r\kappa(ar^2+1)}\,,
\end{align}
\end{subequations}
where $h=h(r)$ and $f=f(r)$. The equation for $h(r)$ can be integrated directly. Inserting its solution into Eq.~\eqref{dfdr}, we find that the system admits the following metric functions as a solution
\begin{subequations}\label{solQBtrivial}
\begin{align}
    h(r)&=\frac{ar^2+1}{br^2+1}\,,\\
    \notag
f(r)&=K\left[\frac{4(ar^2+1)^2(br^2+1)\lambda}{r^4\kappa(b-a)}-\frac{2(ar^2+1)(br^2+ar^2+2)(br^2+1)}{r^4\kappa(b-a)^2}\right]\\
    \notag
    &\quad-\frac{\Lambda(br^2+1)(b^2r^4-6abr^4-3a^2r^4-4br^2-12ar^2-8)}{3(b-a)^3r^4}\\
    &\quad + \frac{c[(br^2+1)(ar^2+1)]^{3/2}}{r^4} + \frac{4(br^2+1)(ar^2+1)(br^2+ar^2+2)}{r^4(b-a)^2}\,,
\end{align}    
\end{subequations}
where $b$ and $c$ are integration constants, such that $a\neq b$. This solution also has a two-dimensional set of fixed points at $r=r_b$ defined as the largest positive root of $f(r_b)=0$. In this case, regularity implies that $\psi\sim\psi+\bar{\beta}$
\begin{align}
    \bar{\beta} = \beta\sqrt{\frac{br_b^2+1}{ar_b^2+1}} \,,
\end{align}
where $\beta$ is given in Eq.~\eqref{betaf-1/f}. Thus, the dependence of the temperature on $b$ factorizes in a very simple way, although it has a nontrivial dependence, as it can be seen from Eq.~\eqref{solQBtrivial}. Additionally, the boundary metric of this solution is still a squashed three-sphere but with a different squashing parameter [cf. Eq.~\eqref{dsbdy}]; that is, 
\begin{align}\label{squashingparameter}
    \mu^2 = \frac{4K\lambda a^2}{\kappa(b-a)} + \frac{\Lambda(3a^2+6ab-b^2)}{3(b-a)^3} + \frac{2a(a+b)(2\kappa-K)}{\kappa(b-a)^2} + c\,a^{3/2}\sqrt{b}\,.
\end{align}

The integration constant $c$ is not independent of $r_b$. 
Indeed, the $r_b\to\infty$ limit is smooth, and the former gets fixed in terms of $a$ and $b$. 
Then, from Eq.~\eqref{squashingparameter}, one observes that the $r_b\to\infty$ limit corresponds to the $\mu^2\to 0$ case, where the boundary metric becomes stretched.

In contrast to the solution in Eq.~\eqref{soldeformedCP2-f-1/f}, the limit $a\to0$ can be taken smoothly in the metric~\eqref{solQBtrivial}. In such a case, the line element takes the form 
\begin{equation}
  \dd s^2=\frac{\dd r^2}{f(r)}+\frac{f(r)}{(1+br^2)}\frac{r^2}{4}\sigma_3^2+\frac{r^2}{4}\dd\Omega^2_{(2)}\,,
\end{equation}
where the metric function that solves the Einstein-Skyrme equations is given by
\begin{align*}\label{fsola0}
    f(r)&=\left(\frac{1+br^2}{r^4}\right)\Bigg[\frac{4K\lambda}{\kappa b}-\frac{2(K-2\kappa)(2+br^2)}{\kappa b^2} -\frac{\Lambda (b^2r^4-4br^2-8)}{3b^3}+c(1+br^2)^{1/2}\Bigg]\,.
\end{align*}
This solution differs from that obtained in Ref.~\cite{Canfora:2013osa}. This can be seen as follows: First, notice that both solutions have the same area coordinate, allowing us to compare them directly from their curvature invariants. Then, one can compute the Kretschmann invariant of both solutions and check that their asymptotic behaviors differ as $r\to\infty$. This proves they are different in a coordinate-invariant way. Additionally, the solution in Eq.~\eqref{fsola0} does not have a smooth $b\to0$ limit when $a=0$, since the Kretschmann behaves as $\mathcal{O}(b^{-4})$ when $\Lambda=0$ and as $\mathcal{O}(b^{-6})$ if $\Lambda\neq0$.  

Direct evaluation of the on-shell action for the solution in Eq.~\eqref{solQBtrivial} yields 
\begin{align*}\notag
I_E& =\pi\bar{\beta} K\lambda\ln\left[\frac{\left(2+r_b^2(a+b)+2\sqrt{(1+ar_b^2)(1+br_b^2)}\right)}{\left(2+r_b^2(a+b)-2\sqrt{(1+ar_b^2)(1+br_b^2)}\right)}\frac{\left(\sqrt{a}-\sqrt{b}\right)^2}{\left(\sqrt{a}+\sqrt{b}\right)^2}\right]\\
&\quad -\frac{\Lambda\kappa\bar{\beta}\pi}{3(a-b)^2}\left[\frac{br_b^4(3a-b)+r_b^2(3a+b)+2}{\sqrt{1+br_b^2}(1+ar_b^2)^{3/2}}+\frac{(b^2-3ab)}{a^{3/2}\sqrt{b}}\right]\\
&\quad +\frac{\bar{\beta}\pi\sqrt{b}}{\sqrt{a}(a-b)^2}\left[4K(a^2\lambda-ab\lambda+b)-\frac{\Lambda\kappa(3a^2-14ab+3b^2)}{3a(a-b)}\right]\\
&\quad +\frac{\bar{\beta}\kappa\pi}{2}\left[(a-3b)c-\frac{16 b^{3/2}}{\sqrt{a}(a-b)^2}\right]\,.
\end{align*}
Similar to the previous case, the logarithmic piece of the partition function is a consequence of the Skyrme term, since it vanishes in the $\lambda\to0$ limit, where the theory reduces to a non-linear sigma model. Additionally, $\beta$ is finite in the $r_b\to\infty$ limit, rendering the on-shell action equal to
\begin{align}
     I_E = \frac{4\pi^2 \kappa}{a}\,.
\end{align}

Considering the line element in Eq.~\eqref{deformedCP2} with the solutions of Eqs.~\eqref{soldeformedCP2-f-1/f} and~\eqref{solQBtrivial}, we find that both the integral of the Gauss-Bonnet and the Chern form are finite. Direct computation of the Euler characteristic of both solutions yields $\chi=2$, differing in one unit with the Euler characteristic of $\mathbb{CP}^2$. This is consistent with the presence of boundaries. 


\section{Gravitational calorons\label{sec:calorons}}

In this section, we construct the gravitational analog of calorons in Einstein-Skyrme theory. These configurations represent instantons periodic in Euclidean time. However, as it has been emphasized, it is possible to construct neither self-dual instantons nor calorons in the usual sense within the Einstein-Skyrme theory since the obvious BPS bound cannot be saturated. Nevertheless, the techniques described previously work also in the case of calorons. To this end, similar to the Euclidean wormhole case, we assume a Skyrme field aligned with the left-invariant forms of $SU(2)$. For the metric, we consider a Euclidean Friedmann-Lemaitre-Robertson-Walker ansatz, which can be obtained from Eq.~\eqref{ansatzWH} by performing a radial diffeomorphism, while relabeling the radial coordinate by a Euclidean time coordinate, i.e. $r\to\tau$; this leads to the following ans\"atze
\begin{align}\label{caloronansatz}
    \diff{s^2} = \diff{\tau^2} + a(\tau)\diff{\Omega_{(3)}^2} \;\;\;\;\; \mbox{and} \;\;\;\;\;  B_\mu = \sigma_\mu^i t_i \,.
\end{align}
Importantly, this choice solves Eq.~\eqref{eomB} automatically, regardless of the form of $a(\tau)$. Inserting the solution of the Skyrme equation into the Einstein-Skyrme field equations~\eqref{eomg}, we obtain 
\begin{subequations}\label{eqs-caloron}
    \begin{align}
        \frac{3}{4}\left(\frac{a'^2}{a^2}+\frac{K\lambda}{\kappa a^2}+\frac{(K-4\kappa)}{\kappa a}\right)+\Lambda&=0\,,\\
        \frac{a''}{a}-\frac{a'^2}{4a^2}+\frac{K-4\kappa}{4\kappa a}-\frac{K\lambda}{4\kappa a^2}+\Lambda&=0\,,
    \end{align}
\end{subequations}
where $a=a(\tau)$ and prime denotes differentiation with respect to $\tau$. If $\Lambda>0$, the system~\eqref{eqs-caloron} is solved by the metric function
\begin{align}\label{caloron}
    a(\tau)=\frac{3(4\kappa-K)}{8\kappa\Lambda} \pm  \frac{\sqrt{9(4\kappa-K)^2-48\kappa K\lambda\Lambda}}{8\kappa\Lambda}\sin\left(\sqrt{\frac{4\Lambda}{3}}\;(\tau-\tau_0)\right)\,,
\end{align}
where $\tau_0$ is an integration constant. This is compact, regular, periodic, and positive definite metric $\forall \tau\in\mathbb{R}$ if and only if $0<K<4\kappa$ and 
\begin{align}
    0<\lambda<\frac{3(4\kappa-K)^2}{16\kappa \Lambda K}\,.
\end{align}
The Euclidean time is identified as $\tau\sim\tau+\beta_\tau$, whose period is $\beta_\tau=\pi\sqrt{3/\Lambda}$. This solution represents a finite-temperature gravitational analog of calorons in Yang-Mills theory. The latter confirms in a quite elegant way the original intuition of Skyrme, that is, the Skyrme term supports smooth regular and topologically non-trivial solitonic solutions. The existence of regular calorons critically relies on the nontriviality of the Skyrme term since, if $\lambda=0$, the solution in Eq.~\eqref{caloron} becomes singular when $\sqrt{4\Lambda/3}\,(\tau-\tau_0)=n\pi/2$ with $n\in\mathbb{N}$. Thus, the Skyrme term is able to support regular calorons which, otherwise, would not exist. The bulk action, the Euler characteristic, and the Chern-Pontryagin index vanish for this configuration. Thus, it represents a ground state with a topologically nontrivial matter configuration since, as in the Euclidean AdS wormhole case, the topological baryonic charge is given by Eq.~\eqref{topologicalcharge1}.

The vanishing cosmological constant case must be analyzed separately since it is not continuously connected with the solution in Eq.~\eqref{caloron}. Taking $\Lambda=0$ in the Einstein-Skyrme system and assuming the ansatz of Eq.~\eqref{caloronansatz}, we find that the solution is given by
\begin{align}\label{caloronL0}
    a(\tau) = \frac{K\lambda}{4\kappa-K} + \frac{(4\kappa-K)(\tau-\tau_0)^2}{4\kappa}\,,
\end{align}
where $a(\tau)>0\,,\forall \tau\in\mathbb{R}$ if and only if $\lambda>0$ and $0<K<4\kappa$. This is a regular positive-definite solution of Einstein-Skyrme equations. This solution represents a regular gravitational instanton whose hypersurfaces of constant $\tau$ are topologically $\mathbb{S}^3$, similar to the case with $\Lambda>0$. It is worth mentioning that regularity is intimately related to the Skyrme coupling since, if $\lambda=0$, there is a curvature singularity at $\tau=\tau_0$. 

\section{Discussion\label{sec:discussion}}

In the present manuscript, we developed an analytic method to construct Euclidean topological solitons in the four-dimensional Einstein-Skyrme theory. The interest of our analysis arises, in part, from the fact that asymptotically AdS Euclidean wormholes provide a natural setup for studying the AdS/CFT correspondence with multiple boundaries. Moreover, from the viewpoint of the semi-classical evaluation of the path integral, regular Euclidean solutions disclose important non-perturbative effects. Here, we have shown that Euclidean-AdS wormholes also exist in the four-dimensional Einstein-Skyrme theory with a negative cosmological constant and positive curvature of its conformal boundaries. Furthermore, the Skyrme matter source possesses a nontrivial topological charge, which is interpreted as the baryonic charge of the configurations. Interestingly enough, the corresponding free energy does not depend on the integration constants, as it usually happens with topological solitons. Additionally, we obtain the holographic stress tensor and $1$-point function of the dual $SU(2)$ current of this configuration and showed that they vanish. This fact allows us to interpret this configuration as a holographic BPS state despite the fact that the corresponding Skyrmions are not BPS solitons in the obvious sense. 

We have also constructed other topologically nontrivial configurations in Einstein-Skyrme theory representing gravitational instantons which are periodic in Euclidean time; these solutions are also known as calorons. Quite remarkably, the regularity of these gravitational calorons is closely related to the Skyrme coupling constant, as in the limit of vanishing Skyrme coupling these calorons become singular. We also studied the free energy as a function of the temperature and found that Hawking-Page phase transitions take place above a critical temperature. The latter depends on a new parameter that appears in the Fubini-Study metric of $\mathbb{CP}^2$ but here is no longer related to the cosmological constant. 

There are many possible extensions of the present results. First of all, it is natural to extend the Euclidean AdS wormhole solution to the $SU(N)$ case. This will provide a natural setup for studying the large $N$ limit of these topological Euclidean solitons in the Einstein-Skyrme theory. Even more, as shown in Ref.~\cite{Concha:2023qcs,Vera:2025qqz}, extending the gauge group to $SU(N)$ allows one to increase the Skyrmions' topological charge, giving access to topologically inequivalent sectors of the theory. Additionally, the relation between the deformed $\mathbb{CP}^2$ space and the double copy in the presence of Skyrmions is certainly relevant for understanding the non-Abelian dual of these configurations. We leave these questions open for future studies.

\begin{acknowledgments}
    The authors thank Giorgos Anastasiou, Felipe D\'iaz, Daniel Flores-Alfonso, and Rodrigo Olea, for insightful comments and discussions. This work is partially supported by Agencia Nacional de Investigación y Desarrollo (ANID), Chile, through Fondecyt Regular grants 1210500, 1230112, 1240043, and 1240048. B.~D. is supported by Becas de Magíster UNAP. 
\end{acknowledgments}

\bibliography{biblio.bib}

\end{document}